\documentclass[dvips]{acta}
\usepackage{supertabular,lscape,epsfig}
\usepackage{amssymb}
\usepackage{amsmath}

\SetPages{0}{0}

\SetVol{59}{2009}

\begin{document}

\begin{Titlepage}
\Title{\bf Rotation of $\delta$~Scuti Stars in the Open Clusters \\
NGC\,1817 and NGC\,7062\footnote{Partly based on observations made 
with the Nordic Optical Telescope, 
operated on the island of La Palma jointly by Denmark, Finland, Iceland, Norway, and 
Sweden, in the Spanish Observatorio del Roque de los Muchachos of the Instituto de 
Astrofisica de Canarias.}
}

\Author{J.~M~o~l~e~n~d~a~-~\.Z~a~k~o~w~i~c~z}{Astronomical Institute, University of  Wroc{\l }aw, Kopernika 11, Wroc{\l }aw, Poland\\
e-mail: molenda@astro.uni.wroc.pl}

\Author{T.~A~r~e~n~t~o~f~t}{Department of Physics and Astronomy, Danish AsteroSeismology Center (DASC), Aarhus University, Ny Munkegade 120, Bldg.\ 1520, 8000 \AA rhus, Denmark\\
e-mail: toar@phys.au.dk}

\Author{S.~F~r~a~n~d~s~e~n}{Department of Physics and Astronomy, Aarhus University, Ny Munkegade 120,
Bldg.\ 1520, 8000 \AA rhus C, Denmark\\
e-mail: sfr@phys.au.dk}

\Author{and F.~G~r~u~n~d~a~h~l}{Department of Physics and Astronomy, Aarhus University, Ny Munkegade 120,
Bldg.\ 1520, 8000 \AA rhus C, Denmark\\
e-mail: fgj@phys.au.dk}

\Received{Month Day, Year}
\end{Titlepage}

\Abstract{We report results of spectroscopic and photometric observations of ten $\delta$~Scuti 
stars and one eclipsing binary in the open cluster NGC\,1817, and of ten $\delta$~Scuti 
stars and two other variables in the open cluster NGC\,7062. For all targets in NGC\,1817 and 
for three targets in NGC\,7062, the radial velocity and projected rotational velocity are 
determined. 
For all stars, the effective temperature and surface gravity is measured.

Two $\delta$~Scuti stars, NGC\,1817~--V1 and NGC\,7062~--V1, and the eclipsing binary, 
NGC\,1817~--V18, are discovered to be single-lined spectroscopic binaries. The 
eclipsing binary $\delta$~Scuti 
star NGC\,1817~--V4 is discovered to be a double-lined spectroscopic binary. 

All $\delta$~Scuti stars which we observed spectroscopically are found to be moderate 
or fast rotators.}{Open clusters: NGC\,1817, NGC\,7062 -- Stars: pulsating: $\delta$~Scuti 
-- Stars: rotation -- Binaries: spectroscopic}

\section{Introduction}

The $\delta$~Scuti--type pulsators are A--F stars with masses between 
1.5 and 2.5$\rm M_{\odot}$. They oscillate mainly in radial and non-radial 
$p$-modes, show light amplitudes at the mmag level and periods of 
the order of hours. Most $\delta$~Scuti stars are multiperiodic which  
makes them promising targets for asteroseismology because every frequency 
carries information about the inner structure of the star and therefore, each one 
provides additional constraints which can be used in tests of stellar models. These 
tests become more stringent if more parameters can be measured independently 
as is the case for clusters where one can assume the same age and metalicity for 
all cluster members in asteroseismic modeling.

A suitable open cluster for a study of $\delta$~Scuti variables has 
an age of 0.3--1.0 Gyr and a distance of 1--2 kpc. This ensures a 
convenient angular size of the cluster and allows precise photometry of 
$\delta$~Scuti stars, which in this case are found among the cluster's 
brightest members. NGC\,1817 and NGC\,7062 meet these requirements. They 
are rich in $\delta$~Scuti stars (see Arentoft et al.~2005 and Freyhammer 
et al.~2001) and their angular diameter allows finding the targets in a 
field of view of a few arc minutes on the sky. 

In this paper, we derive the effective temperature and surface gravity of
ten $\delta$~Scuti stars and one eclipsing binary in NGC\,1817, and of ten
$\delta$~Scuti stars and two other variables in NGC\,7062. For all
11 stars in NGC\,1817 and for three $\delta$~Scuti stars in NGC\,7062, we 
measure the radial velocity and the projected rotational velocity. We plan to 
use this last parameter, $v\sin i$, as a constraint for target selection in
the future multi-site asteroseismic campaigns on $\delta$~Scuti stars in these clusters. 

This paper is organized as follows. In Section 2, we give an account of
our observations and reductions. In Section 3, we determine $T_{\rm eff}$ 
and $\log g$ of the program stars. In Section 4, we derive the radial velocity
and the projected rotational velocity. Section 5 contains a summary.

\section{Observations and Reductions}

\subsection{Photometry of NGC\,7062}

Str{\"o}mgren photometry for NGC\,7062 and standard stars 
were collected at the 2.5-m Nordic Optical Telescope, La Palma, Spain, on
three nights in August 2004. Since the third night, 14 August 2004, 
was non-photometric, it was excluded from the analysis, which is then based 
on 12 and 13 August 2004. The log of our observations is given in Table 1.

The standard stars were adopted from Olsen (1983, 1984) who provides $uvby-\beta$ 
magnitudes for stars faint enough to be observed with the NOT. The telescope, 
however, still had to be defocused in order to avoid very short integration times.
The adopted exposure times were 5--40 seconds and, to decrease the dead-time 
between observations, only a small window of the CCD was read out. 

During the observations of NGC\,7062, the telescope was in focus and full 
CCD was read out. The exposure times were 60\,s in the $y$ filter, 
90\,s, in $b$, 240\,s, in $v$, 1200\,s, in $u$, 120\,s, in the narrow $\beta$-filter, and 
400\,s, in the wide. 

\MakeTable{lrrc|lrrc}{12.5cm}{Log of the Str{\"o}mgren observations of NGC\,7062: 
the numbers of $uvby$ and $\beta$ sequences and the range in airmass.}
{\hline\noalign{\smallskip}
Date and object  & $uvby$ & $\beta$ & Airmass &Date and object  & $uvby$ & $\beta$ & Airmass\\
\noalign{\smallskip}
\hline
\noalign{\smallskip}
\bf{12 august 2004}&    &   &                    & \bf{13 August 2004} \\ 
\noalign{\smallskip}
Standard stars     & 18 & 8 &  1.01 -- 2.15      & Standard stars     & 11 & 10 & 1.01 -- 1.85       \\
NGC\,7062          & 5  & 4 &  1.05 -- 1.42      & NGC\,7062          & 5  &  4 & 1.22 -- 1.58       \\
\noalign{\smallskip}\hline
}

The observations of the standard stars were reduced with aperture photometry. We used the 
curves of growth to make sure that all photons were included, and we estimated the
sky level using the surrounding background. 
The instrumental $uvby$-magnitudes for the standard stars were used to 
derive transformation equations of the form given in Grundahl et al.~(2002), 

\begin{equation}
y_{obs}=V_{std}+\alpha_{y}(v-y)+\beta_{y}(X-1)+\gamma_{y}T+\delta_{y}
\end{equation}
\begin{equation}
b_{obs}=b_{std}+\alpha_{b}(v-y)+\beta_{b}(X-1)+\gamma_{b}T+\delta_{b}
\end{equation}
\begin{equation}
v_{obs}=v_{std}+\alpha_{v}(v-y)+\beta_{v}(X-1)+\gamma_{v}T+\delta_{v}
\end{equation}
\begin{equation}
u_{obs}=u_{std}+\alpha_{u}(v-y)+\beta_{u}(X-1)+\gamma_{u}T+\delta_{u},
\end{equation}
where $X$ is the airmass, $T$, the time of the exposure, and $(v-y)$, 
the color index in the standard system. 
The transformation equations were derived on a nightly basis. 

The extinction coefficients for the narrow and wide $H_{\beta}$ filters were 
computed from the observations of standard stars for each night separately, and
used to calculate the average extinction coefficient, $k_{\beta}$. Then, the 
total fluxes measured for standard stars were expressed in the units of flux per 
second, converted to magnitudes and corrected for extinction according to the
following equations
\begin{equation}
H_{n,0} = H_{n,obs} - k_{\beta}X
\end{equation}
\begin{equation}
H_{w,0} = H_{w,obs} - k_{\beta}X
\end{equation}
where X is the airmass, and $H_{n,obs}$ and $H_{w,obs}$, magnitudes in the narrow 
and wide $H_{\beta}$ filter, respectively. Finally, the transformation equation
was derived,
\begin{equation}
\beta_{std} = \alpha\beta_{0} + \delta
\end{equation}
where $\beta_{0} = H_{w,0}-H_{n,0}$. 

In Table 2, we give the standard deviation of the residuals resulting from 
subtracting the magnitudes of the standard stars transformed with 
equations 1--4, $m_{trans}$, from the standard magnitudes, $m_{std}$, and
the standard deviations of the residuals of $\beta_{std}- \beta_{trans}$.
The values show that our transformations yield magnitudes which are 
accurate typically to better than 1\%. Furthermore, in the residuals no trends with 
airmass, color or time are present. 

\MakeTable{lrrrrr}{12.5cm}{Standard deviations of the residuals, 
$m_{\rm std} - m_{\rm transf}$, computed for $-uvby-\beta$ observations of 
standard stars (in mmag).}
{\hline\noalign{\smallskip}
Date  & $\sigma _y$ & $\sigma _b$ & $\sigma _v$ & $\sigma _u$ & $\sigma _{\beta}$  \\
\noalign{\smallskip}\hline\noalign{\smallskip}
12 August 2004 & 4.4 & 3.0 & 7.4 &  7.7 & 10.8 \\ 
13 August 2004 & 6.0 & 5.4 & 5.1 & 13.7 & 9.4 \\ 
\noalign{\smallskip}\hline
}

The photometric reductions of the images of NGC\,7062 were done with
the software package MOMF (Kjeldsen \& Frandsen 1992) which uses a combined
PSF and aperture photometry. In this package the size of the aperture is specified by 
the user. We used MOMF for detecting stars in the images and for deriving the 
stellar fluxes. When computing the fluxes for stars in NGC\,7062, we used smaller 
apertures than those used for standard stars because of the crowding in the field of the cluster.
For isolated cluster stars, we used the
curves of growth to verify that the chosen apertures were large enough to include
all the photons coming from the star and, at the same time, small enough to protect against
undesirable crowding effects.

We transformed the instrumental magnitudes to the standard system using the 
equations $1-4$ and $7$. This transformation was done iteratively on a nightly basis.
For each star, the first estimate of $(v-y)_{std}$ was calculated from all measurements
on a given night from sum equations, e.g., for the $y$-filter
\begin{equation}
\sum_{n}{y_{obs}}=nV_{std}+n\alpha_{y}(v-y)+\beta_{y}\sum_{n}{(X-1)}+
\gamma_{y}\sum_{n}{T}+n\delta_{y}
\end{equation}
which was solved for $y_{std}$ and $v_{std}$.
These values were used to transform the individual measurements 
in each filter so that outlying measurements could be discarded before calculating
the nightly mean for a given star. The individual measurements in each filter
were used for calculating the uncertainty of the mean. This procedure was then 
repeated for the other filters. 
The standard $uvby-\beta$ magnitudes were  
calculated as weighted means of the measurements from the first and the second 
night. The mean differences between the magnitudes
measured on the two nights were smaller than 0.01 mag in all cases and typically
equal to few mmag for stars brighter that V=15 mag.

In Table 3, we give the equatorial coordinates, 
the standard $uvby-\beta$ magnitudes, and the $m_1$ and $c_1$ Str\"omgren indices 
for each star. We use values greater than 50 to code those $\beta$ measurements which 
were measured on one night only and are uncertain. The code 
'$0.000$' is used for those uncertainties
of individual $uvby$ magnitudes which were derived from one or two individual
points, which is not enough to determine the standard deviation. The code '99.999' is used
for magnitudes which were not measured. The full table is available in the electronic form 
from the Acta Astronomica Archive (see the cover page). A sample, containing the heading and
the $uvby-\beta$ magnitudes from the first five rows and the last row, is printed below. 

\MakeTable{rccccccr}{12.5cm}{A sample of the $uvby-\beta$ photometry of stars in NGC\,7062.}
{\hline\noalign{\smallskip}
 ID& $\alpha _{\rm 2000}$ & $\delta _{\rm 2000}$ & y & b & v & u & $\beta$ \hspace*{6pt}\\
\noalign{\smallskip}\hline\noalign{\smallskip}
  1 &21:23:19.4&  46:20:23.4&     16.066&   16.937&  18.131& 19.832& 2.621\\
  2 &21:23:30.4&  46:20:23.3&     18.564&   19.772&  21.244& 23.198& 2.649\\
  3 &21:23:24.8&  46:20:24.0&     18.228&   18.951&  19.925& 21.220& 2.479\\
  4 &21:23:30.1&  46:20:23.8&     14.740&   16.075&  18.143& 20.640& 2.617\\
  5 &21:23:44.0&  46:20:22.6&     20.329&   21.054&  22.037& 22.654&51.287\\
\dotfill&\dotfill&\dotfill&\dotfill&\dotfill&\dotfill&\dotfill&\dotfill\\
904 &21 23 34.30&  46 26 40.90&     99.999 &  99.999&  21.900& 99.999&99.999\\
\noalign{\smallskip}\hline
}

\begin{figure}[htb]
\includegraphics{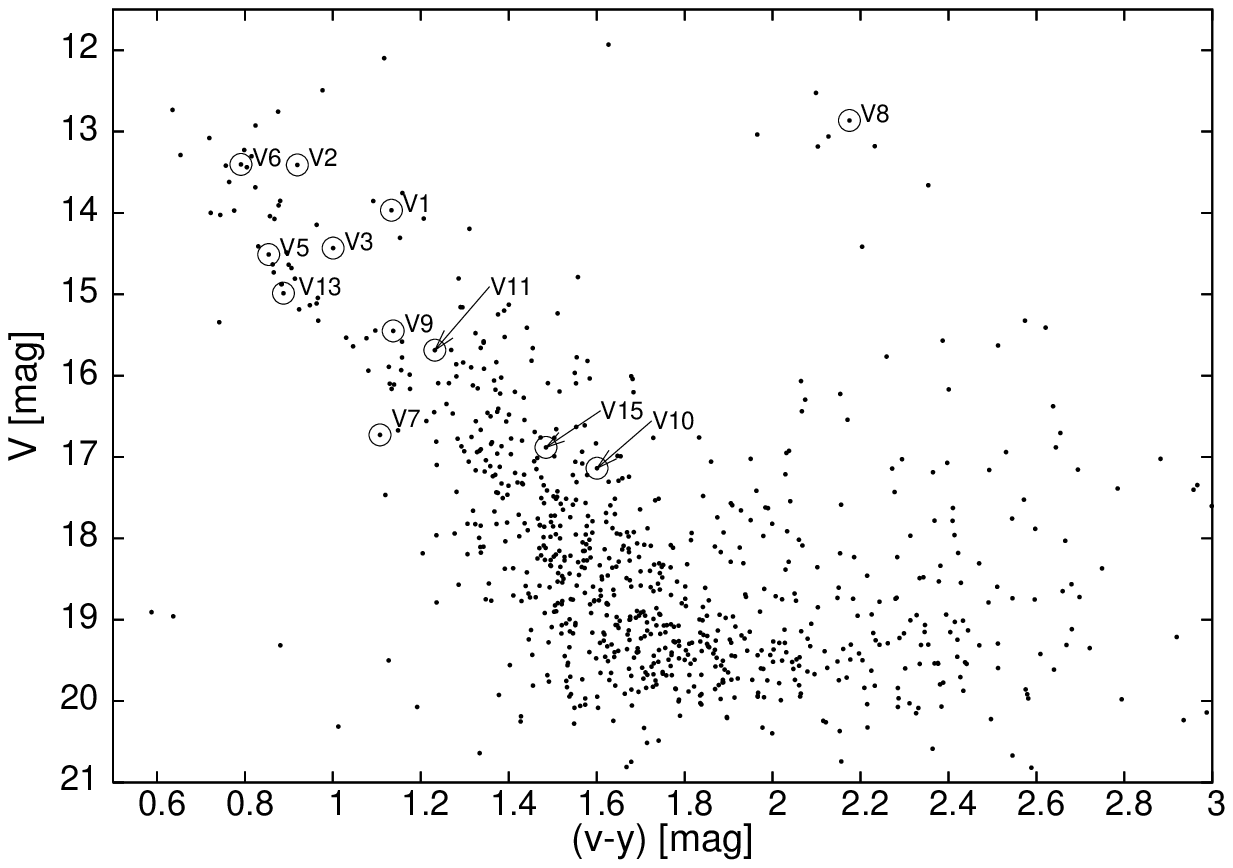}
\FigCap{Color - magnitude diagram for stars observed in the open 
cluster NGC\,7062 and listed in Table 3. Variable stars are indicated with circles. 
}
\end{figure}

In Fig.\ 1, we plot the color-magnitude diagram for stars in the field of 
NGC 7062. We use circles to indicate variable stars which we discuss in Sect.\ 3. 

As the precision of our photometry drops for stars 
fainter than $V=18$ mag, for computing the color excess of NGC\,7062
we used only main-sequence stars brighter than this limit. We de-reddened 
each star by means of the method of Crawford (1979) and computed a 
weighed mean $E(B-V)= 0.43 \pm0.05$ mag. 
Then, we compared the $(b-y)-c_1$ diagrams constructed for NGC\,7062 and
for M\,11 (=NGC\,6705). For the latter cluster, whose $E(B-V) = 0.428\pm0.027$ 
(see Sung et al.~1999), we used existing but unpublished Str\"omgren photometry obtained
by F.G. We found that the difference of the color excess between these two 
clusters is smaller than 0.04 mag and hence the two values agree with each 
other to within the one standard deviation. Therefore, we adopt $E(B-V)= 0.43$
as the color excess of NGC\,7062.

A similar comparison of $(v-y)-V$ and $c_1-V$ diagrams constructed for these two 
clusters (not shown) yields a difference in the distance modulus between
M\,11 and NGC\,7062 equal to $-$0.2 mag, which places NGC\,7062 0.2 mag closer than M\,11. 
Sung et al.~(1999) provide a distance modulus of M11, $V_0 - M_V = 11.55 \pm 0.1$,
which corresponds to an apparent distance modulus of 12.92 for that cluster
and thus results in 12.72 for NGC7062. The last value is in a good agreement with $12.76\pm0.4$
given by Freyhammer et al.~ (2001). 

\subsection{Spectroscopy of NGC\,1817 and NGC\,7062}

The spectroscopic observations were carried out at the Nordic Optical 
Telescope by S.\ F.\ (13--14 August 2004), and by T.\ A.\ and 
J.\ M.--\.Z.\ (20--22 November 2004). The Andalucia Faint Object 
Spectrograph and Camera, 
ALFOSC, was used on both observing runs. The ALFOSC setup consisted 
of grism 9 working in the echelle mode and grism 11 working as a 
cross-disperser. A 2\,048$\times$2\,048 px NIMO Back Illuminated CCD42-40 was
used as the detector. The total wavelength range of the 
spectrograms was 3900--10350\AA \, and their resolution ranged from 
4500 at 5000\AA~to 3900 at 7500\AA. 

The list of targets included eleven stars in NGC\,1817 and three in 
NGC\,7062. The typical exposure time was 1800\,s, 
which resulted in a signal-to-noise ratio of around 60 for most stars. 
The data were reduced with the IRAF\footnote{IRAF is 
distributed by the National Optical Astronomy Observatories, which are 
operated by the Association of Universities for Research in Astronomy, 
Inc., under cooperative agreement with the National Science Foundation.}
software, and the spectra were extracted with the {\sf apall} task also 
provided by IRAF. We give the log of our spectroscopic observations in Table 6.

\section{Effective Temperature and Surface Gravity}

\subsection{From Str\"omgren indices}

We derived the effective temperature and surface gravity of stars 
in NGC\,7062 and NGC\,1817 using Str\"omgren indices from our Table 3 and from
Balaguer-N\'u\~nez et al. (2004), respectively. For the computations, we used the 
calibration of Napiwotzki et al.~(1993) and the code kindly provided by the authors.
We give the values of $T_{\rm 
eff}$ and $\log g$ in Table 4 (columns headed `observations')
together with standard deviations calculated from observational 
errors of the Str\"omgren indices $\beta$ and $c_0$. The large dispersion of 
$\sigma _{T_{\rm eff}}$ for stars in NGC\,1817 results from the very
high dispersion of $\sigma _\beta$ given by Balaguer-N\'u\~nez et al.~(2004), which 
can vary by a factor of 20 from star to star.

\MakeTable{lrcll|r@{$\pm$}lr@{$\pm$}l|rr}{12.5cm}{Variable stars in 
NGC\,1817 and NGC\,7062 from Arentoft et al.~(2005) and Freyhammer et al.~(2001), 
respectively. BN04 refers to the identification number from Balaguer-N\'u\~nez 
et al.~(2004), ID, to the number from Table 3. The effective temperature 
and gravity computed from Str\"omgren indices and model atmospheres are given in the 
columns headed 'observations' and 'models'.
}
{\hline\noalign{\smallskip}
\multicolumn{10}{c}{NGC\,1817}\\
\noalign{\smallskip}\hline\noalign{\smallskip}
star  &  BN04 & V & $\alpha _{\rm 2000}$ & $\delta _{\rm 2000}$&
      $T_{\rm eff}$ & $\sigma$ & $\log g$ & $\sigma$ & 
      $T_{\rm eff}$ & $\log g$ \\
&&&&&\multicolumn{4}{|c|}{observations}&\multicolumn{2}{c}{models}\\
\noalign{\smallskip}\hline\noalign{\smallskip}
V1 & 167& 13.53 &05 12 42.8&+16 41 43& 7991 &             434 & 3.96 & 0.06 & 7750 & 4.0\\
V2 & 154& 12.88 &05 12 40.8&+16 42 00& 7298 &             317 & 3.50 & 0.06 & 7250 & 3.5\\
V3 & 184& 14.42 &05 12 37.4&+16 42 31& 7962 &             707 & 4.49 & 0.02 & 7750 & 4.5\\
V4 &7615& 12.62 &05 12 32.3&+16 44 52& 7095 &             264 & 3.66 & 0.15 & 7000 & 3.5\\
V6 & 156& 12.95 &05 12 33.1&+16 41 50& 7648 &             700 & 4.14 & 0.05 & 8000 & 4.5\\
V7 &7632& 13.73 &05 12 40.1&+16 46 07& 7521 &             527 & 3.94 & 0.15 & 7500 & 4.0\\
V8 & 423& 14.37 &05 12 33.7&+16 43 22& 6968 &             816 & 4.07 & 0.17 & 7250 & 4.5\\
V9 &7294& 13.18 &05 12 24.6&+16 43 32& 6893 &\hspace{2pt}  38 & 3.57 & 0.16 & 7250 & 4.0\\
V11&7097& 14.32 &05 12 30.4&+16 41 29& 8075 &             589 & 4.65 & 0.09 & 8250 & 5.0\\
V12&7105& 14.71 &05 12 27.9&+16 40 06& 7333 &             174 & 4.60 & 0.13 & 7250 & 4.5\\
V18&7145& 14.24 &05 12 30.0&+16 37 30& 7091 &\hspace{2pt}  98 & 4.12 & 0.11 & 7000 & 4.0\\
\noalign{\smallskip}\hline\noalign{\smallskip}
\multicolumn{10}{c}{NGC\,7062}\\
\noalign{\smallskip}\hline\noalign{\smallskip}
star & ID & V & $\alpha _{\rm 2000}$ & $\delta _{\rm 2000}$&
      $T_{\rm eff}$ & $\sigma$ & $\log g$ & $\sigma$ & 
      $T_{\rm eff}$ & $\log g$ \\
&&&&&\multicolumn{4}{|c|}{observations}&\multicolumn{2}{c}{models}\\
\noalign{\smallskip}\hline\noalign{\smallskip}
V1 &357 &13.97 &21 23 29.8&+46 22 38&  7749 &             185 & 4.20 & 0.02 & 7750 & 4.5\\
V2 &184 &13.41 &21 23 30.5&+46 21 38&  8074 &             193 & 3.54 & 0.02 & 8000 & 3.5\\
V3 &594 &14.43 &21 23 35.8&+46 24 10&  7945 &             345 & 3.91 & 0.03 & 8000 & 4.0\\
V5 &728 &14.51 &21 23 21.7&+46 25 12&  8218 &             282 & 3.91 & 0.04 & 8000 & 4.0\\
V6 &396 &13.40 &21 23 21.5&+46 22 59&  8500 &             347 & 4.05 & 0.04 & 8250 & 4.0\\
V7 &836 &16.73 &21 23 46.3&+46 26 00&  9577 &             218 & 3.54 & 0.40 & 9750 & 4.5\\
V8 & 69 &12.86 &21 23 13.6&+46 20 51& 14812 &             235 & 2.49 & 0.02 &14000 & 2.0\\
V9 &562 &15.45 &21 23 40.1&+46 23 56&  7694 &             490 & 4.40 & 0.06 & 7500 & 4.5\\
V10&328 &17.14 &21 23 22.9&+46 22 25&  7906 &             594 & 4.19 & 0.09 & 8000 & 4.5\\
V11&270 &15.69 &21 23 33.4&+46 22 07&  7526 &             509 & 4.60 & 0.07 & 7250 & 4.5\\
V13&145 &14.99 &21 23 18.5&+46 21 23&  8235 &             265 & 3.93 & 0.12 & 8250 & 4.0\\
V15&395 &16.88 &21 23 23.8&+46 22 58&  7035 &             587 & 5.21 & 0.17 & 6750 & 5.0\\
\noalign{\smallskip}\hline
}

Comparing the values of $\log g$ given by Balaguer-N\'u\~nez et al.~(2004) for stars in NGC 1817
with those listed in Table 4, we find a satisfactory agreement. Unfortunately, such an
agreement does not exist for the effective temperatures and absolute magnitudes, as the stars 
discussed by Balaguer-N\'u\~nez et al.~(2004) are found to be cooler and brighter than in this 
paper. We discuss these discrepancies in more detail in the Appendix.

\subsection{From 2MASS photometry}

We compared the values of $T_{\rm eff}$ derived from Str\"omgren indices 
with those resulting from 2MASS indices $(V-J)$, $(V-H)$ and
$(V-K)$. For these computations we used the calibration of Kinman \& Castelli (2002). 
For NGC\,1817, we 
adopted $E(B-V)= 0.28$ from Balaguer-N\'u\~nez et al.~(2004), for NGC\,7062, $E(B-V)= 
0.430$ from this paper. We de-reddened the 2MASS magnitudes
using the $A(\lambda)/E(B-V)$ ratios provided by Cutri et al.~(2003) and  
transformed the de-reddened magnitudes to the Bessell \& Brett (1988) 
homogenized system. 
For NGC\,1817 we adopted $\rm [Fe/H] = -0.30$, which is a mean of 
the $\rm [Fe/H]$ determinations from the literature that range from $-0.42$ (Taylor 
2001) to $-0.26$ (Twarog et al.~1997). For NGC\,7062, we used $\rm [Fe/H] = -0.35$ as 
derived by
Peniche et al.~(1990) from photometry. The values of $\log g$ which are required
by the calibration of  Kinman \& Castelli (2002) were adopted from Table 4
(the columns headed 'observations.')

The effective temperatures derived from the three color indices  are given in Table 5. 
This table does not list NGC\,7062--V7 and --V8, which are too hot for the calibration,
NGC\,7062--V11 and --V13, for which 2MASS magnitudes are of low quality 
(the 2MASS quality flag, Qflg, set to 'E'), and NGC\,7062--V15, which is too 
faint for the 2MASS Catalogue. The standard deviations of $T_{\rm eff}$ include 
photometric errors and the uncertainty of $\log g$ from Table 4. 

\MakeTable{l|c|l l l}{12.5cm}{Atmospheric parameters determined from $JHK$ indices for 
stars in NGC\,1817 and NGC\,7062.}
{\hline\noalign{\smallskip}
\multicolumn{5}{c}{NGC\,1817}\\
\noalign{\smallskip}\hline\noalign{\smallskip}

star& 2MASS ID &$ T_{\rm eff}^{V-J} \pm \sigma$& 
                $ T_{\rm eff}^{V-H} \pm \sigma$& 
                $ T_{\rm eff}^{V-K} \pm \sigma$\\
\noalign{\smallskip}\hline\noalign{\smallskip}

V1 & 05124282+1641432 & 8118$\pm$61 & 8044$\pm$58 & 8037$\pm$65 \\
V2 & 05124081+1642003 & 7970$\pm$50 & 7858$\pm$45 & 7784$\pm$55 \\
V3 & 05123739+1642308 & 7982$\pm$26 & 7839$\pm$41 & 7722$\pm$60 \\
V4 & 05123227+1644518 & 7956$\pm$49 & 7815$\pm$47 & 7813$\pm$54 \\
V6 & 05123307+1641503 & 7963$\pm$30 & 7827$\pm$40 & 7721$\pm$47 \\
V7 & 05124010+1646072 & 8224$\pm$80 & 8105$\pm$68 & 8213$\pm$92 \\
V8 & 05123370+1643220 & 7917$\pm$37 & 7717$\pm$63 & 7658$\pm$71 \\
V9 & 05122460+1643324 & 7882$\pm$39 & 7664$\pm$47 & 7595$\pm$47 \\
V11& 05123036+1641285 & 8091$\pm$47 & 7929$\pm$43 & 7782$\pm$54 \\
V12& 05122787+1640062 & 7959$\pm$26 & 7658$\pm$57 & 7500$\pm$69 \\
V18& 05122997+1637296 & 7961$\pm$37 & 7814$\pm$39 & 7687$\pm$48 \\
\noalign{\smallskip}\hline\noalign{\smallskip}
\multicolumn{5}{c}{NGC\,7062}\\
\noalign{\smallskip}\hline\noalign{\smallskip}
star& 2MASS ID & $ T_{\rm eff}^{V-J}\pm \sigma$& 
               $ T_{\rm eff}^{V-H} \pm \sigma$& 
               $ T_{\rm eff}^{V-K} \pm \sigma$\\
\noalign{\smallskip}\hline\noalign{\smallskip}
V1 & 21232977+4622378 & $7988\pm \,\,\,\,25$  & $7834\pm \,\,\,\,30 $  & $7783\pm \,\,\,\,35 $ \\
V2 & 21233057+4621376 & $7916\pm \,\,\,\,27$  & $7757\pm \,\,\,\,28 $  & $7720\pm \,\,\,\,30 $ \\
V3 & 21233583+4624102 & $7993\pm \,\,\,\,33$  & $7807\pm \,\,\,\,32 $  & $7743\pm \,\,\,\,41 $ \\
V5 & 21232168+4625115 & $7903\pm \,\,\,\,13$  & $7769\pm \,\,\,\,32 $  & $7650\pm \,\,\,\,56 $ \\
V6 & 21232154+4622592 & $8126\pm \,\,\,\,49$  & $8032\pm \,\,\,\,42 $  & $7971\pm \,\,\,\,43 $ \\
V9 & 21234015+4623555 & $7839\pm \,\,\,\,41$  & $7225\pm 144$          & $6704\pm 226$\\
V10& 21232292+4622246 & $8179\pm \,\,\,\,79$  & $7974\pm \,\,\,\,75$   & $7890\pm 144$\\
\noalign{\smallskip}\hline
}

\begin{figure}[htb]
\includegraphics{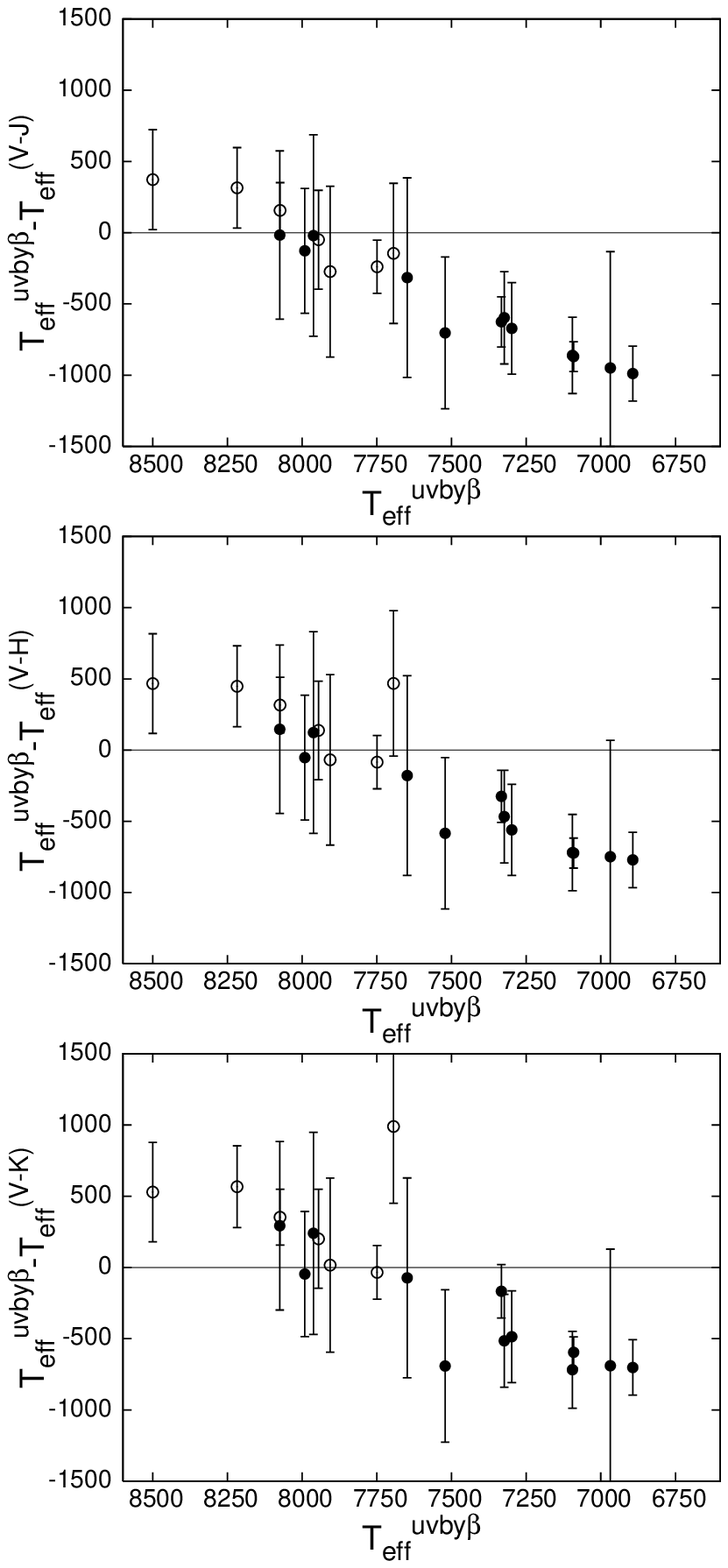}
\FigCap{Differences between $T_{\rm eff}$ computed from Str\"omgren and
2MASS indices. Dots indicate stars in NGC\,1817, open circles, stars in NGC\,7062.
}
\end{figure}

In Fig.\ 2, we show the differences between $T_{\rm eff}$ obtained from the
calibration of Napiwotzki et al.~(1993) and those from the
calibration of Kinman \& Castelli (2002). The differences are positive for
stars hotter than about 8000\,K and negative, for the cooler.
This trend is present for $T_{\rm eff}$ derived from 
each 2MASS photometric index and is most clearly visible 
in $(V-J)$, in which the maximum difference in $T_{\rm eff}$ computed for 
NGC\,1817--V9 reaches 1000 K.
The existence of this trend was not expected as no such feature is present 
in the comparison of $T_{\rm eff}$ derived by Kinman \& Castelli (2002) 
from their calibration and by means of other methods.

\subsection{From model atmospheres}

We derived the effective temperature and surface gravity of the 
program stars using Kurucz ODFNEW model atmospheres\footnote{available at 
http://kurucz.harvard.edu/} computed for $\rm [M/H] = 0.0$ and $-0.5$, 
with the step in $T_{\rm eff}$ and $\log g$ equal to 250\,K and 0.5 
dex, respectively (Castelli \& Kurucz 2004, Castelli \& Kurucz 2006). 
In Table 4 (columns headed `models'), we give $T_{\rm eff}$ and $\log g$ 
of the model atmosphere of  $\rm [M/H] = 0.0$, for which the synthetic Str\"omgren 
indices are closest to the observed ones. For models computed for $\rm [M/H] = -0.5$, 
either the resulting values of $T_{\rm eff}$ and $\log g$ were the same or no 
match was found. 

\begin{figure}[htb]
\includegraphics{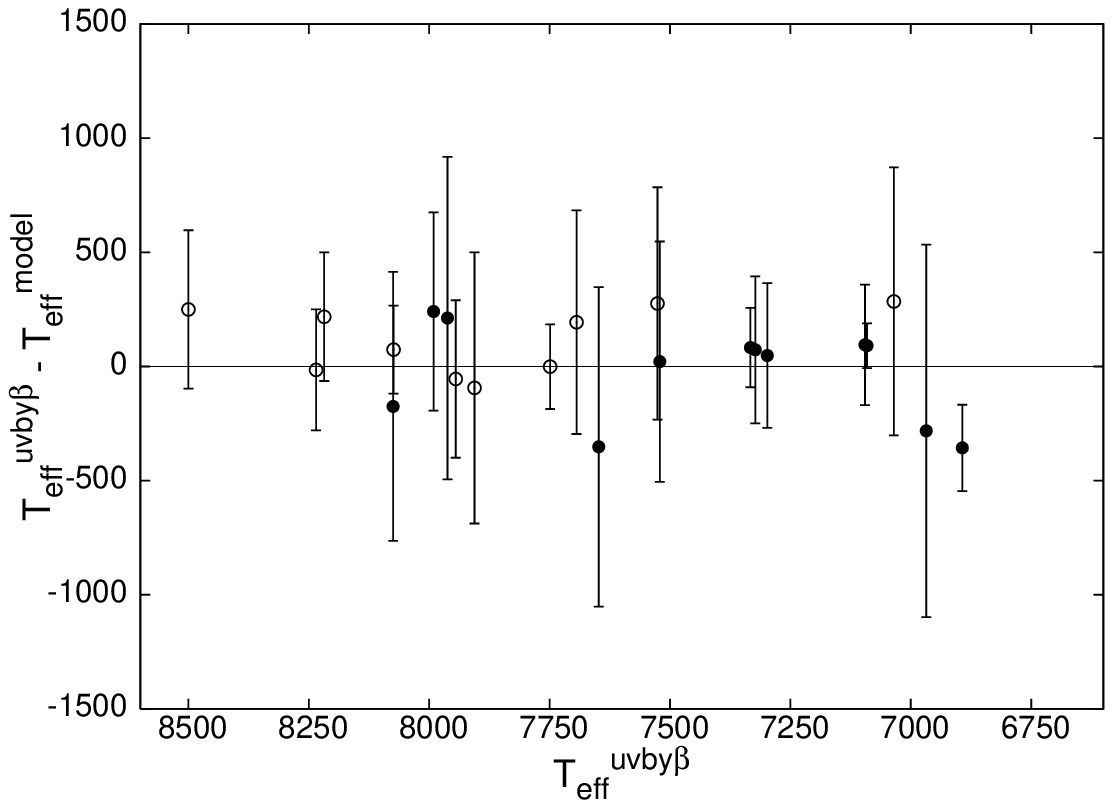}
\FigCap{Differences between $T_{\rm eff}$ computed from Str\"omgren
indices and model atmospheres.
Dots indicate stars in NGC\,1817, open circles, stars in NGC\,7062.}
\end{figure}

In Fig.\ 3, we plot the differences 
between $T_{\rm eff}$ derived from the calibration of Napiwotzki et al.~(1993) 
and those derived from the model atmospheres. The values agree 
well to within one standard deviation in all but one case 
and no trend is present. Having reached this consistency, we use the 
$T_{\rm eff}$ and $\log g$ obtained from the calibration of Napiwotzki et al.\
(1993) in the next steps of our analysis.

\begin{figure}[htb]
\includegraphics{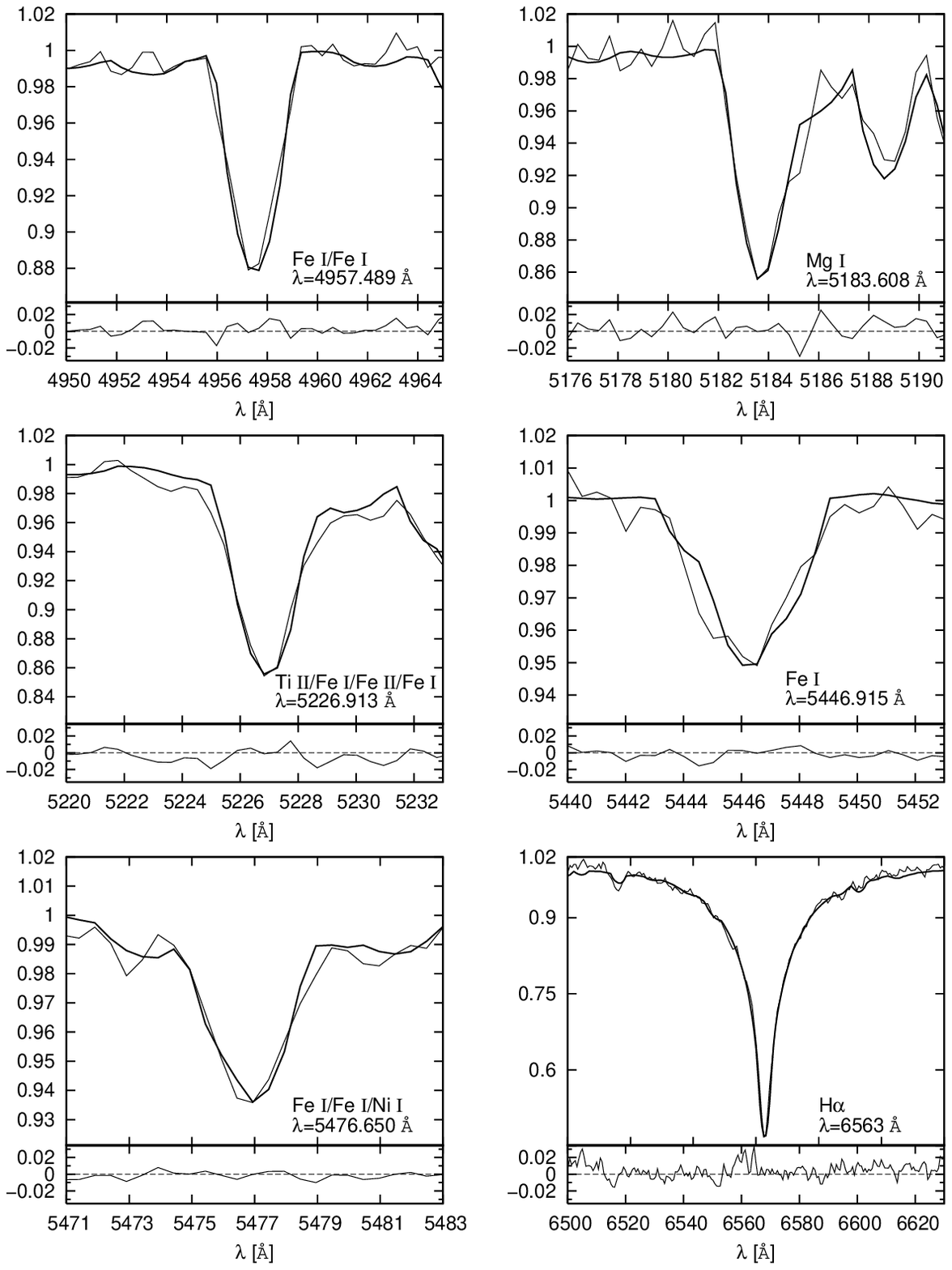}
\FigCap{The observed (thin line) and fitted spectrum (thick line), and the
residuals of the fit calculated for six lines in NGC\,1817--V2.} 
\end{figure}

\section{Radial Velocity and the Projected Rotational Velocity}

The observed spectra with the ALFOSC instrument at the NOT have
rather low resolution for detailed studies of radial and rotational
velocities. But even with an instrumental width of the order 50 km/s
one can still classify stars as fast or slow rotators or as binary stars.

We measured the radial velocity, $RV$, and $v\sin i$ of the program stars using two methods. 
First, we used the cross-correlation method and the {\sf fxcor} task provided 
by IRAF. We obtained $RV$ measurements in each order of the spectrograms and 
then computed the weighted mean $RV.$ As templates, we used synthetic spectra 
computed from Kurucz ODFNEW model atmospheres.
We used the ATLAS9 and SYNTHE (Sbordone et al.~2004, Sbordone 2005)
software to compute the dedicated model atmosphere and the synthetic
template spectrum for each program star for $T_{\rm eff}$ and $\log g$
adopted from Table 4 (columns headed 'observations')
and $\rm [Fe/H]$ equal to $-0.30$ and $-0.35$ for NGC\,1817 and
NGC\,7062, respectively. 

The projected rotational velocity was derived
by comparing the observed spectrum with the synthetic one. The synthetic
spectrum was rotationally broadened in a wide range of $v\sin i$ according
to the formula provided by Gray (1992). Since the spectrograms measured by us
have relatively low S/N ratios and most of our targets turned out to be 
fast rotators, the precise measurements of $RV$ were difficult; the 
spectral lines were blurred and
merged which caused problems with fitting the continuum and for determining 
whether the spectra correspond to single or multiple stars.
For these reasons, we used mainly strong hydrogen lines but
the metallic lines from Table 3 of Rasmussen et al.~(2002)
were used whenever possible. In Fig.\ 4, we show fragments of the observed 
spectrum of NGC\,1817--V2, one of our brightest targets, over-plotted with the
fitted synthetic spectrum. At the bottom of each panel of the figure
we show the residuals of the fit.

In the second method, we calculated broadening functions as described by Ruci\'nski 
(1999), again using the synthetic spectra as templates. The broadening 
functions were computed individually for each echelle order and then a weighted 
average was formed. The weights were based on the deviation 
of each broadening function from the average. Finally, the rotational velocity 
and the barycentric radial velocity were derived 
by fitting a convolution of a Gaussian instrumental broadening and a rotational 
profile to the observed broadening function. The baseline and the amplitudes 
were kept fixed. For the binary stars, or for stars that might be binaries 
according to the shape of the broadening function, we fit two components. This
was difficult because the components overlap and, as the broadening 
function has some noise peaks, these can be misinterpreted as a stellar component.
We kept in mind, however, that if a binary star spectrum is assumed to come from a 
single star, it is possible to measure an artificially large $v \sin i$.
In conclusion, we found that a significant fraction of the stars have broad 
profiles and that they clearly rotate at rather high rates.

\begin{figure}[htb]
\includegraphics{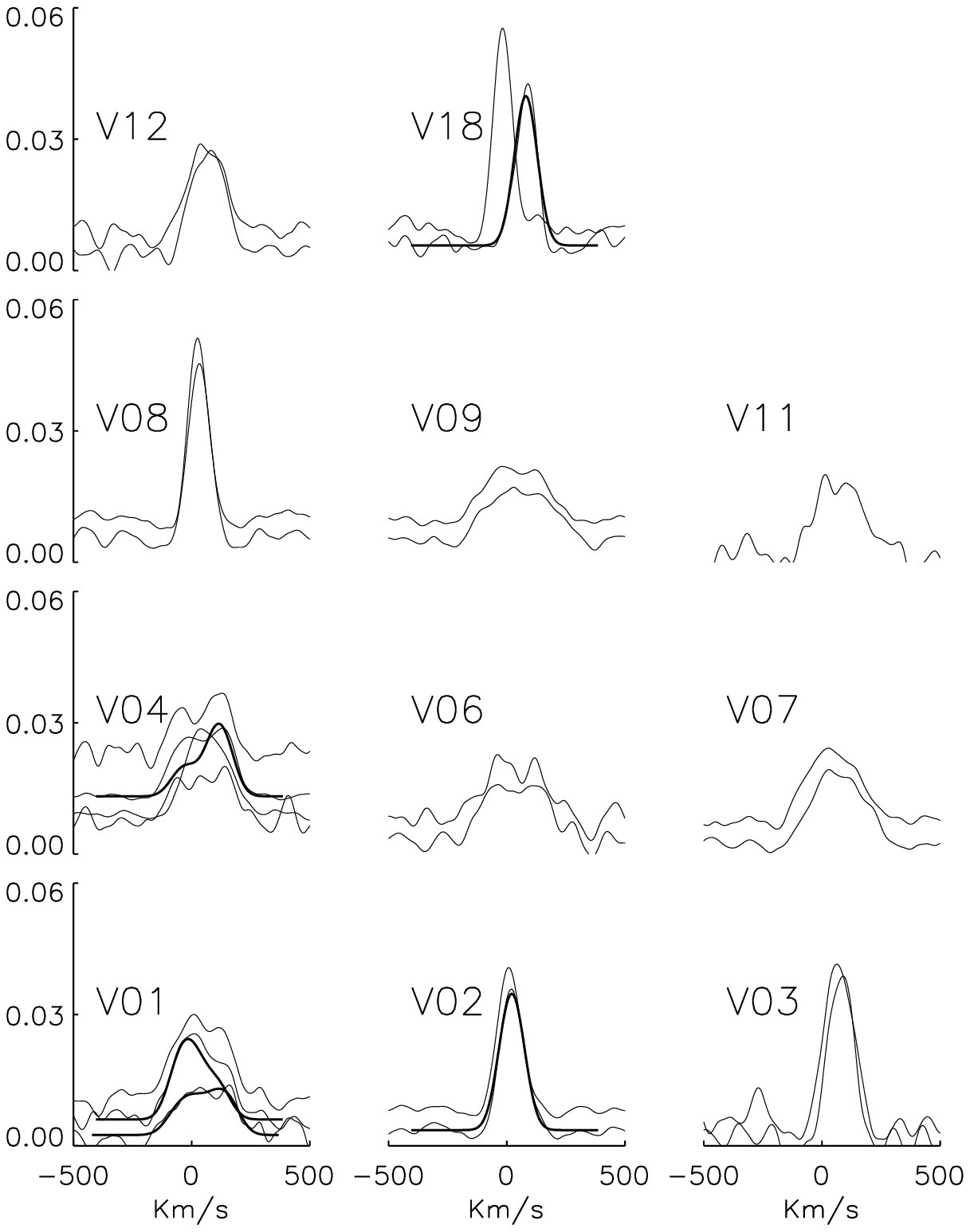}
\FigCap{The broadening functions for $\delta$ Sct stars and the SB1 eclipsing binary V18 in NGC 1817.}
\end{figure}

In Fig.\ 5, we show the broadening functions calculated for ten $\delta$~Scuti stars  in NGC\,1817
and the eclipsing binary, NGC\,1817--V18. The broadening functions for stars in NGC\,7062 (not shown) look similar.
Each panel of Fig.\ 5 is labeled with the number of the variable and shows several 
broadening functions shifted slightly upwards one with respect to another 
for the sake of clarity of the diagram.  For four stars, we use thick lines 
to show the quality of the fitted curves. The broadening functions of the 
spectroscopic binaries, V01, V04 and V18, are clearly variable.

\MakeTable{lrccr|lrccr}{12.5cm}{Logbook of observations
of NGC\,1817 and NGC\,7062. The columns contain star designations
from Arentoft et al.~(2005) and Freyhammer et al.~(2001) for NGC\,1817 and NGC\,7062,
respectively, the projected rotational velocity, $v\sin i$,
the Heliocentric Julian Date of the mid-exposure, the exposure time in seconds,
and the radial velocity.}
{\hline\noalign{\smallskip}
star&$v\sin i$ & HJD    &$\rm t_{exp}$ & $V_r \pm \sigma$ & star&$v\sin i$ & HJD    &$\rm t_{exp}$ & $V_r \pm \sigma$\\
    &[km/s]    &-2450000& [s]          & [km/s]           &     &[km/s]    &-2450000&[s]           & [km/s]          \\
\noalign{\smallskip}\hline\noalign{\smallskip}
\multicolumn{5}{c|}{NGC\,1817}&\multicolumn{5}{c}{NGC\,7062}\\
\noalign{\smallskip}\hline\noalign{\smallskip}
V1 &194&3232.7382 & 1200  &  98.6          13.3     &V1 & 84&3231.4900 & 1800 & -28.9  2.6\\
   &   &3330.5414 & 1800  &  58.0 \hspace*{1pt} 9.3 &   &   &3231.5141 & 1800 & -14.9  1.1\\
   &   &3330.5671 & 1800  &  63.0 \hspace*{1pt} 6.8 &   &   &3232.5166 & 1800 &  -2.1  1.1\\
\multicolumn{5}{c|}{\dotfill}&\multicolumn{5}{c}{\dotfill}\\
V2 & 60&3331.6481 & 1800  &  32.5 \hspace*{1pt} 2.7 &V3 &175&3231.5744 & 1800 &  20.7  6.8\\
   &   &3331.6721 & 1800  &  27.3 \hspace*{1pt} 4.4 &   &   &3231.5988 & 1800 &  25.1  6.1\\
\multicolumn{5}{c|}{\dotfill}& \multicolumn{5}{c}{\dotfill}\\
V3 & 76&3331.7363 & 1800  &  86.4 \hspace*{1pt} 4.2 &V5 & 90&3231.6288 & 1800 &  -1.0  2.8\\
   &   &3331.7602 & 1800  &  75.7 \hspace*{1pt} 4.5 &   &   &3231.6533 & 1800 & -11.3  2.9\\
\multicolumn{5}{c|}{\dotfill}&             \\
V4 &167&3330.4666 & 1800  &  58.7 13.6 \\
   &   &3331.5022 & 1800  &  57.1 10.7 \\
   &   &3331.7056 & 1800  &  89.3 \hspace*{1pt} 7.1 \\
   &   &3332.5656 & 1800  &  67.7 \hspace*{1pt} 7.8\\
\multicolumn{5}{c|}{\dotfill}\\
V6 &225&3332.4422 & 1800  &  68.5 14.5 \\
   &   &3332.4760 & 1800  &  57.3 12.9\\
\multicolumn{5}{c|}{\dotfill}\\
V7 &177&3330.6640 & 1800  &  77.6 \hspace*{1pt} 4.4\\
   &   &3330.6881 & 1800  &  66.4 \hspace*{1pt} 6.0\\
\multicolumn{5}{c|}{\dotfill}\\
V8 & 50&3330.6031 & 1800  &  47.4 \hspace*{1pt} 3.1\\
   &   &3330.6277 & 1800  &  42.6 \hspace*{1pt} 2.0\\
\multicolumn{5}{c|}{\dotfill}\\
V9 &222&3332.5102 & 1800  &  54.9 \hspace*{1pt} 5.6\\
   &   &3332.5341 & 1800  &  44.3 \hspace*{1pt} 4.0\\
\multicolumn{5}{c|}{\dotfill}\\
V11&164&3332.6331 & 2400  &  82.0 \hspace*{1pt} 7.2\\
\multicolumn{5}{c|}{\dotfill}\\
V12&119&3332.6680 & 1800  &  74.3 \hspace*{1pt} 3.7\\
   &   &3332.6919 & 1800  &  63.8 \hspace*{1pt} 4.0\\
\multicolumn{5}{c|}{\dotfill}\\
V18& 45&3330.7226 & 1800  &  99.4 \hspace*{1pt} 5.5\\
   &   &3332.5986 & 1800  &   1.4 \hspace*{1pt} 3.8\\
\noalign{\smallskip}\hline
}

In Table 6, we list the $RV$ computed with the use of the methods described above.
The values range from 20 to 80 km/s; this  
dispersion is much larger than for open clusters in general but still not
large enough that the membership to the cluster is ruled out for any of the stars.
We suspect that the instrumental drift of the order 10 km/s might account for
some of this dispersion.

\begin{figure}[htb]
\includegraphics{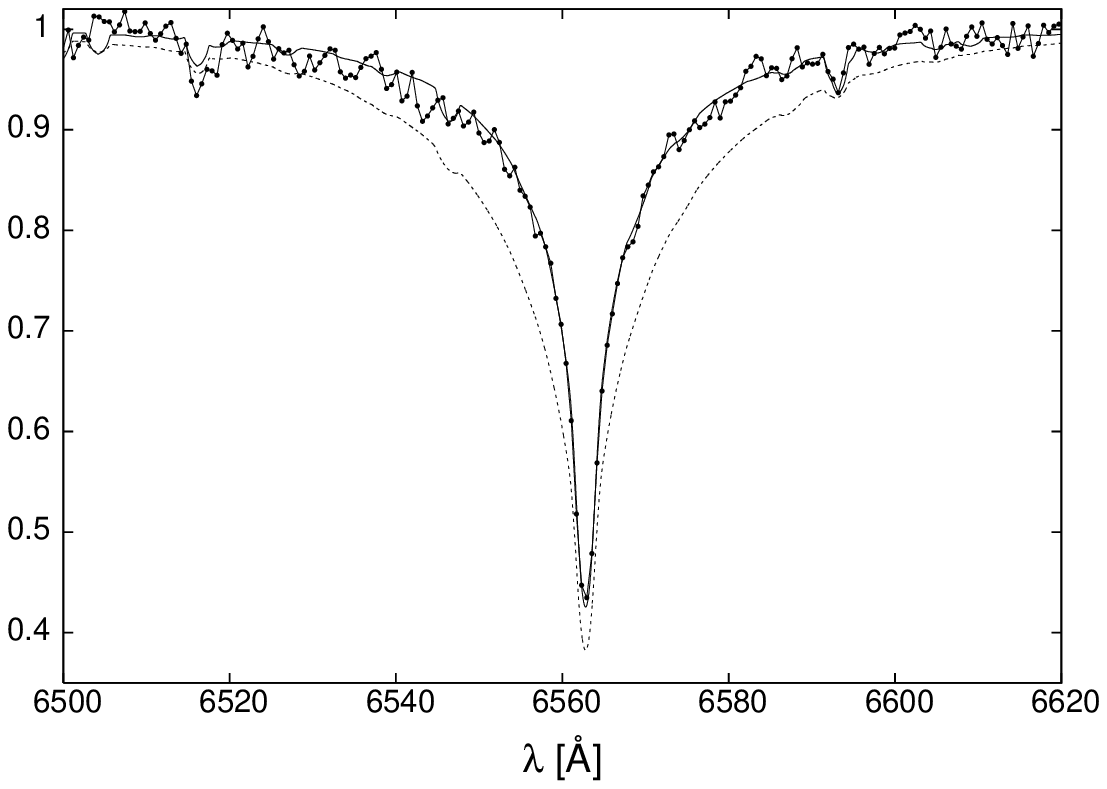}
\FigCap{NGC\,1817--V8: the observed H$\alpha$ line over-plotted with two 
rotationally broadened synthetic H$\alpha$ lines computed for the same $\log g$ 
and two different $T_{\rm eff}$ (see the text).}
\end{figure}

Four stars from the studied sample show significant variability in radial velocity.
Three of them, NGC\,1817--V1, --V18, and NGC\,7062--V1, we classify as 
single-lined spectroscopic binaries, and one, NGC\,1817--V4, which was
discovered to be an eclipsing binary by Arentoft et al.~(2005),  as a new 
spectroscopic double-lined binary. The latter classification was done on the 
basis of the shape of the broadening function. A few of the remaining stars 
also show indications of variability, but the uncertainties of the $RV$ 
measurements are large and do not allow drawing firm conclusions.

In the second column of Table 6, we give the $v\sin i$ calculated as a mean of
the values obtained from all examined spectral 
lines. The measured values range from 45 to 225 km/s for NGC\,1817, and from 84
to 175 km/s for NGC\,7062. The mean $v\sin i$ of $\delta$~Scuti stars 
in these two clusters are equal to 145$\pm$21 and 116$\pm$29 km/s, 
respectively. These numbers fall close to the median $v\sin i$ of $\delta$ 
Scuti stars in the Hyades cluster and in NGC\,6134 (see Table 7
of Rasmussen et al.~2002). This adds NGC\,1817 and NGC\,7062 to the list 
of clusters in which $\delta$~Scuti stars rotate fast. We note, however, 
that for NGC\,7062 the mean $v\sin i$ was calculated from three stars only.

Finally, in Fig.\ 6, we show H$\alpha$ line of NGC\,1817--V8  
for which the discrepancies between the effective temperature
determined from the Str\"omgren and 2MASS indices are of the highest. The measured 
spectrum is over-plotted with two synthetic H$\alpha$ lines. The first, 
plotted with a solid line, computed for a model atmosphere with $T_{\rm 
eff} = 6968$ and $\log g = 4.07$ derived from Str\"omgren indices, and the second, plotted 
with a dashed line, for a model atmosphere with the same $\log g$ 
but $T_{\rm eff} = 7250$ derived from $(V-J)$ index. Both 
synthetic spectra are rotationally broadened to 50 km/s. As can be seen, 
the synthetic spectrum computed for $T_{\rm eff}$ 
derived from $(V-J)$ index is substantially different from the observed one. For 
other stars with large differences in $T_{\rm eff}$ derived from Str\"omgren indices
and 2MASS photometry, the discrepancies are similar. Therefore, we conclude that 
for stars discussed in this paper, atmospheric parameters obtained from Str\"omgren 
indices are more adequate, as they provide a better agreement between the observed and 
the computed spectra.

\section{Summary}

We reported results of spectroscopic and photometric observations of stars in NGC\,1817 
and NGC\,7062, which were intended as preparation for a future multi-site 
asteroseismic campaign on $\delta$ Scuti stars in these clusters. 
The spectrograms were used for determination of the radial velocity and the projected 
rotational velocity of the program stars by means of two different methods.

NGC\,1817--V1, --V18, and NGC\,7062--V1 are discovered to be single-lined spectroscopic 
binaries, NGC\,1817--V4, to be a double-lined spectroscopic binary, and 
all the observed stars are found to be moderate or fast rotators. The latter makes them 
challenging targets for asteroseismology as fast rotators are more complicated 
to model (see, e.g., Kjeldsen et al.\ 1998), and their frequency spectrum is more
difficult to interpret. We note, however, that there are successful attempts of asteroseismic 
analysis of rapidly rotating $\delta$~Scuti stars (see, e.g., Michel et al.~1999).

\Acknow{This publication makes use of data products from the Two Micron All Sky 
Survey, which is a joint project of the University of Massachusetts and 
the Infrared Processing and Analysis Center/California Institute of 
Technology, funded by the National Aeronautics and Space Administration 
and the National Science Foundation.

The data presented here have been taken using ALFOSC, which is owned by 
the Instituto de Astrofisica de Andalucia (IAA) and operated at the Nordic 
Optical Telescope under agreement between IAA and the NBIfAFG of the 
Astronomical Observatory of Copenhagen.

We made use of the SAO/NASA Astrophysics Data System (ADS) and the Asiago 
database on Photometric Systems (ADPS).

J.M.-\.Z.\ and T.A.\ thank the Danish Natural Science Research Council for 
financial support.
J.M.-\.Z.\ acknowledges a partial financial support from the 
MNiSW grant N203 014 31/2650.}

\newpage
\centerline {\vspace{0.5cm}{\bf Appendix I}}

\MakeTable{r|rrrrr|rrrrr}{12.5cm}{$T_{\rm eff}$, $R/\rm R_{\odot}$ and $M_V$ of four stars
in NGC\,1817 derived by Balaguer-N\'u\~nez et al.~(2004) and in this paper,
the absolute visual magnitude $M_V^{\log L}$ which results from 
the total luminosity of the star, and the difference $\Delta M_V = M_V^{\log L} - M_V.$
}
{\hline\noalign{\smallskip}
star& $T_{\rm eff}$ & $R/\rm R_{\odot}$ & $M_V$ & $M_V^{\log L}$ & 
      $\Delta M_V$ &
      $T_{\rm eff}$ & $R/\rm R_{\odot}$ & $M_V$ & $M_V^{\log L}$ & 
      $\Delta M_V$ \\
    &\multicolumn{5}{c|}{Balaguer-N\'u\~nez et al.~(2004)}
    &\multicolumn{5}{c}{this paper}\\
\noalign{\smallskip}\hline\noalign{\smallskip}
V1  & 7794 & 2.55 & 1.77 & 1.38 & -0.39 & 7991 & 1.84 & 2.04 & 1.99 & -0.05\\
V2  & 7105 & 4.61 & 0.57 & 0.49 & -0.08 & 7298 & 3.63 & 0.84 & 0.89 &  0.05\\
V3  & 7731 & 1.37 & 3.30 & 2.76 & -0.54 & 7962 & 1.01 & 3.49 & 3.31 & -0.18\\
V4  & 6917 & 3.38 & 1.37 & 1.29 & -0.08 & 7095 & 2.88 & 1.48 & 1.52 &  0.04\\
\noalign{\smallskip}\hline
}

Balaguer-N\'u\~nez et al.\ (2004) compute atmospheric parameters
of selected stars in NGC\,1817 adopting $\rm [Fe/H] = -0.34$ and using 
the grid of Moon \& Dworetsky (1954) which they modify so that it is sensitive to the 
metalicity (unpublished). They derive the stellar effective temperature from 
the fit of the star's flux to the synthetic spectra in four different bands (V-JHK)
(Masana, private communication, see also Masana el al.~2006). 

We find that the values of $\log g$ given by Balaguer-N\'u\~nez et al.~(2004) agree 
well with those computed in this paper, but the effective temperatures are 
cooler and the absolute magnitudes, brighter. We illustrate these discrepancies in 
the top panel of Fig.\ 7 where we use circles for $T_{\rm eff}$ and $M_V$ 
from Balaguer-N\'u\~nez et al.~(2004) and dots, for this paper. Lines show mean 
$T_{\rm eff}$--$M_V$ relations for dwarfs and giants
from Lang (1992). Arrows join respective pairs of stars. The relevant values of 
$T_{\rm eff}$ and $M_V$ are given in Table 7.

The discrepancies are opposite to what might be expected for metal-poor 
stars whose $T_{\rm eff}$ is determined from the $\beta$ index. We illustrate this 
in the bottom panel of Fig.\ 7, where we plot the $T_{\rm eff}$ -- $\beta$ 
relation for Kurucz ODFNEW model atmospheres computed for $\log g$ = 4.0, and 
$\rm [M/H] = 0.0$ (dashed line) and $-0.5$ (solid line). According to these models, 
of two stars that have the same $\beta$ and $\log g$, the metal-deficient one should 
be hotter. As an example, we show that a star with $\log g = 4.0$, $\beta = 
2.835$ and $\rm [M/H] = 0.0$ should have $T_{\rm eff} =$ 7742\,K (dashed 
arrow) while a star with the same $\log g$ and $\beta$ but $\rm [M/H] = -0.5$
should have $T_{\rm eff} =$ 7795\,K, (solid arrow). As Balaguer-N\'u\~nez et 
al.~(2004) do not use the $\beta$ index 
for the determination of $T_{\rm eff}$, the noticed discrepancies must result 
from differences in the calibrations of $T_{\rm eff}$ in Str\"omgren and JHK indices.

\begin{figure}[htb]
\includegraphics{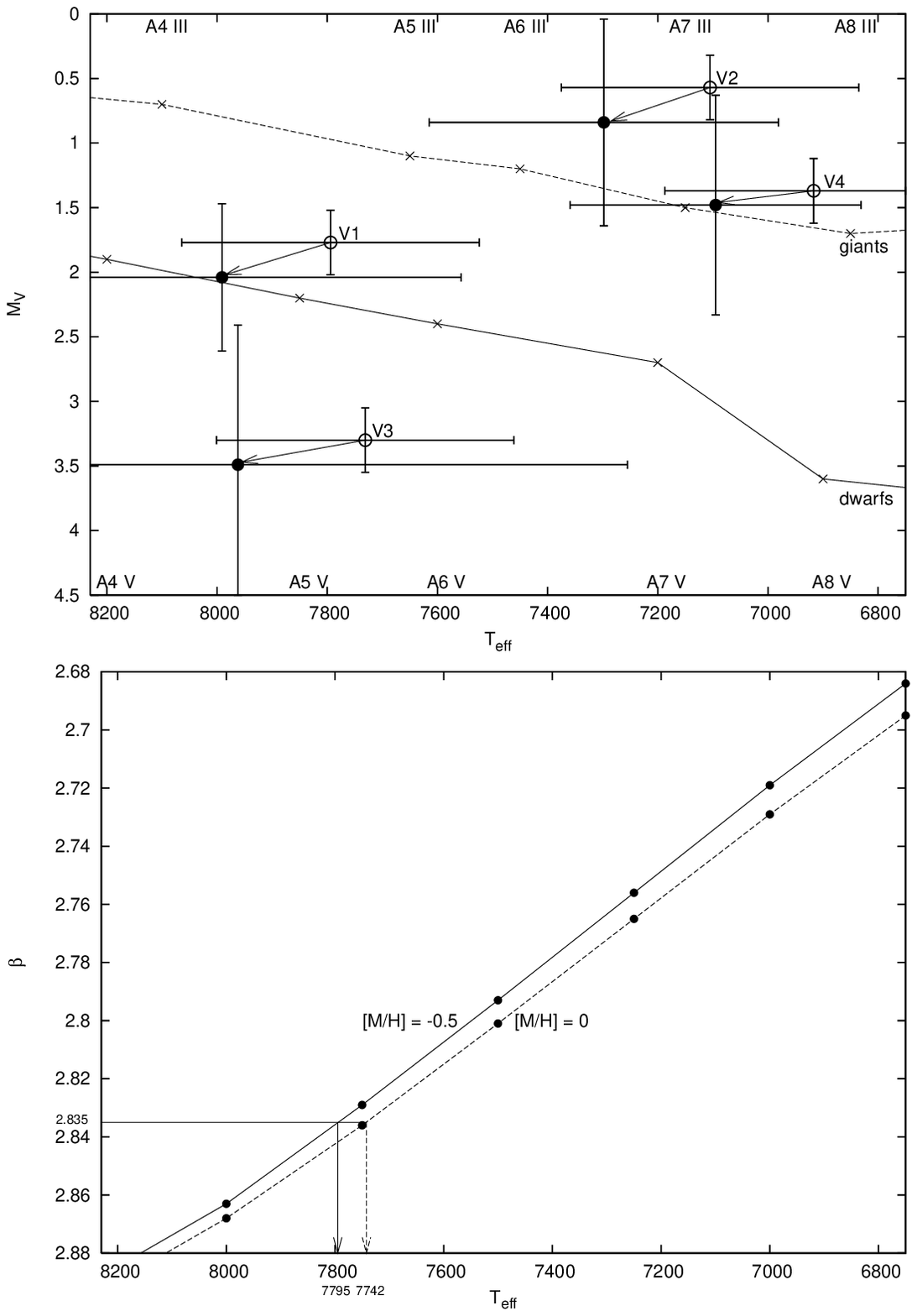}
\FigCap{{\it Top:} $T_{\rm eff}$--$M_V$ diagram for stars in NGC\,1817
discussed in this paper and in Balaguer-N\'u\~nez et al.~(2004).
{\it Bottom:}
$T_{\rm eff}$--$\beta$ relation for stars of different metalicities.
Arrows show that for a fixed $\log g$ and $\beta$ a metal-poor star is
hotter than a star of solar metalicity.}
\end{figure}

Considering the values of $M_V$ listed by Balaguer-N\'u\~nez et al.~(2004), 
we find that in some cases they are significantly different from $M_V^{\log L}$,
the absolute magnitude which results from the total luminosity of the star
\begin{equation}
M_V^{\log L} = M_{\rm bol} - BC 
\end{equation}
where
\begin{equation}
M_{\rm bol} = -2.5\log L/\rm L_{\odot} + M_{\rm bol,\odot}
\end{equation}
and
\begin{equation}
L = 4\pi R^2 \sigma T_{\rm eff}^4
\end{equation}

In Table 7, we calculate the differences between $M_V^{\log L}$ and $M_V$
using $T_{\rm eff}$ and $R/\rm R_{\odot}$ from Table 7 of Balaguer-N\'u\~nez 
et al.~(2004) and the values derived in this paper. We used the bolometric 
correction, BC, from Flower (1996), the solar effective temperature, $T_{\rm eff, 
\odot} = 5\,778$K, and the solar bolometric absolute magnitude, $M_{\rm bol, 
\odot} = 4.75$ mag.  We find that in all cases the consistency between 
$M_V^{\log L}$ and $M_V$ is better for the calibration of Napiwotzki et al.~(1993).
Therefore, we suspect that the large values of $\Delta M_V$ found for Balaguer-N\'u\~nez 
et al.~(2004) may indicate some problems with consistency of their calibration.


\begin{references}
\refitem{Arentoft, T., Bouzid, M.Y., Sterken, C., Freyhammer, L.M., and Frandsen, S.} {2005} {PASP} {117} {601}

\refitem{Balaguer-N\'u\~nez L., Jordi C., Galadi-Enriquez D., and Masana E.} {2004} {Astron.\ Astrophys.} {426} {827}

\refitem{Bessell, M.S., and Brett, J.M.} {1988} {PASP} {100} {1134}

\refitem{Castelli, F., Kurucz, R.L.} {2004} {Proc.\ IAU Symp.\ No 210, {\it Modelling of Stellar Atmospheres}, \rm eds.\ N. Piskunov et al.} {~} {eprint arXiv:astro-ph/0405087} 

\refitem{Castelli, F., Kurucz, R.L.} {2006} {Astron.\ Astrophys.} {454} {333}

\refitem{Crawford D.L.} {1979} {Astron.\ J.} {84} {1858}

\refitem{Cutri R.M.\ et al.} {2003} {The 2MASS All-Sky Catalog of Point Sources, University of Massachusetts and Infrared Processing and Analysis Center IPAC/California Institute of Technology}{~}{~}

\refitem{Flower P.J.} {1996} {Astrophys.\ J.} {469} {355}

\refitem{Freyhammer, L.M., Arentoft, T., and Sterken, C.} {2001} {Astron.\ Astrophys.} {368} {580}

\refitem{Gray, D.F.} {1992} {\it The Observation and Analysis of Stellar Photospheres} {~} {Cambridge University Press}

\refitem{Grundahl, F., Stetson, P.B., and Andersen, M.I.} {2002} {Astron.\ Astrophys.} {395} {481}

\refitem{Kinman, T., and Castelli, F.} {2002} {Astron.\ Astrophys.} {391} {1039}

\refitem{Kjeldsen, H., Arentoft, T., Bedding, T., Christensen--Dalsgaard, J., Frandsen, S., and Thompson, M.J.} {1998} {{\it Structure and Dynamics of the Interior of the Sun and Sun-like Stars} SOHO 6/GONG 98, ESASP} {418} {385}

\refitem{Kjeldsen, H., and Frandsen, S.} {1992} {PASP} {104} {413}

\refitem{Lang, K.R.} {1992} {Astrophysical Data I. Planets and Stars, Springer-Verlag Berlin Heidelberg New York}{~}{~}

\refitem{Masana, E., Jordi, C., and Ribas, I.} {2006} {Astron.\ Astrophys.} {450} {735}

\refitem{Michel, E., Hern\'andez, M.M., Houdek, G., Goupil, M.J., Lebreton, Y., P\'erez Hern\'andez, F., Baglin, A., Belmonte, J.A., and Soufi, F.} {1999} {Astron.\ Astrophys.} {342} {153}

\refitem{Moon, T.T., and Dworetsky, M.M.} {1985} {MNRAS} {217} {305}

\refitem{Napiwotzki, R., Sch\"onberner, D., and Wenske, V.} {1993} {Astron.\ Astrophys.} {268} {653}

\refitem{Olsen, E.H.} {1983} {Astron.\ Astrophys.\ Suppl.\ Ser.} {54} {55}

\refitem{Olsen, E.H.} {1984} {Astron.\ Astrophys.\ Suppl.\ Ser.} {57} {443}

\refitem{Peniche, R., Pe\~na, J.H., Diaz-Martinez, S.H., and Gomez, T.} {1990} {Rev.\ Mex.\ Astron.\ Astrofis.} {20} {127}

\refitem{Rasmussen, M.B., Bruntt, H. Frandsen, S., Paunzen, E., and Maitzen, H.M.} {2002} {Astron.\ Astrophys.} {390} {109}

\refitem{Ruci\'nski S.M.} {1999} {IAU Coll. 170, "Precise Stellar Radial Velocities", Eds.\ Hearnshaw and Scarfe, ASP Conf.\ Ser.} {185} {82}

\refitem{Sbordone, L.} {2005} {Mem.\ Soc.\ Astron.\ Ital.\ Suppl.} {8} {61}

\refitem{Sbordone, L., Bonifacio, P., Castelli, F., and Kurucz, R.L.} {2004} {Mem.\ Soc.\ Astron.\ Ital.\ Suppl.} {5} {93}

\refitem{Sung, H., Bessell, M.S., Lee, H.-W., Kang, Y.H., and Lee, S.-W.} {1999} {MNRAS} {310} {982}

\refitem{Taylor, B.J.} {2001} {Astron.\ Astrophys.} {377} {473}

\refitem{Twarog, B.A., Ashman, K.M., and Anthony-Twarog, B.J.} {1997} {Astron.\ J.} {114} {2556}

\end{references}
\end{document}